**Title Page**

Classification

    Major category: Physical Sciences

    Minor category: Applied Physical Sciences

# Local Structure Order Assisted Two-step Crystal Nucleation in Polyethylene

Xiaoliang Tang[1], Junsheng Yang[1,2], Tingyu Xu[1], Fucheng Tian[1], Chun Xie[1], Liangbin Li[1*]

[1]*National Synchrotron Radiation Lab and CAS Key Laboratory of Soft Matter Chemistry, University of Science and Technology of China, Hefei, China*

[2]*Computational Physics Key Laboratory of Sichuan Province, Yibin University, Yibin, China*

[*] Correspondence author: lbli@ustc.edu.cn
Author contributions: X. T, J. Y, T. X, C. X and F. T designed research; X. T, J. Y and T. X performed research; X. T and L. L contributed new analytic tools; X. T, F. T, T. X analyzed data; X. T and L. L wrote the paper.

**Abstract**

Homogeneous nucleation process of polyethylene (PE) is studied with full-atom molecular dynamic simulation. To account the complex shape with low symmetry and the peculiar intra-chain conformational order of polymer, we introduce a shape descriptor $O_{CB}$ coupling conformational order and inter-chain rotational symmetry, which is able to differentiate hexagonal and orthorhombic clusters from melt. With the shape descriptor $O_{CB}$, we find that coupling between conformational and inter-chain rotational orderings results in the formation of hexagonal clusters first, which is dynamic in nature. Whilst nucleation of orthorhombic structure occurs inside of hexagonal clusters later, which proceeds via the coalescence of neighboring hexagonal clusters rather than standard stepwise growth process. This demonstrates that nucleation of PE crystal is a two-step process with the assistance of $O_{CB}$ order, which is different from early models for polymer crystallization but similar with that proposed for spherical 'atoms' like colloid and metal.

**Significance Statement**

By introducing a shape descriptor $O_{CB}$ that couples intra-chain conformational order and inter-chain rotational order, we successfully differentiate local structures with hexagonal and orthorhombic symmetries and observe $O_{CB}$ order assisted two-step nucleation process in polyethylene crystallization. $O_{CB}$ order is demonstrated to promote the transformation from flexible chains to conformational ordered segments, which is the most peculiar and critical step in polymer crystallization. The shape descriptor $O_{CB}$ may be universal on differentiating local orders in polymer or systems with connectivity.

\body

**Introduction**

Crystal nucleation from supercooled liquid is a fundamental phase transition (1) phenomenon universal to materials as well as biological systems. With density as the single order parameter, classical nucleation theory (CNT) (2) and density functional theories (DFT) (3) provide a framework of one-step liquid-solid transition to depict nucleation successfully on a qualitative level, while it is hard to test their predictions such as on nucleation rate at different conditions quantitatively (4–8) and the molecular details of nucleation still remain elusive. Approaching the molecular mechanism of nucleation, two-step nucleation models, involving in either density (5, 9, 10) or bond-orientational order fluctuation (6, 11–13) prior to crystallization, have been proposed. Although their molecular pathways are not the same, both scenarios emphasize the importance of precursor on nucleation. The existence of precursor during nucleation has been widely reported by computer simulations and experiments on spherical "atoms" like colloid and metal (6, 14–17) as well as complex molecules like protein and synthetic polymers (5, 18–21), suggesting that two-step nucleation process may indeed be a general mechanism of nucleation. Nevertheless, current discussion on bond-orientational order assisted nucleation is mainly restricted in spherical atom systems like colloid and metal with high symmetry, which is rarely mentioned in the nucleation of complex molecular systems like polymer (19, 22, 23). For particles with arbitrary shapes or polymer with complex structures, a local structure order independent from density order may be defined by a shape descriptor with specific mathematic fingerprint.

Two-step or multi-step nucleation scenarios with conformational order or density fluctuation (3, 24–27) have also been proposed to challenge the standard Hoffman-Lauritzen (HL) polymer crystallization model (28). Crystallization of synthetic and natural polymers shares the common nucleation mechanism of small molecules but yet has its own peculiar features, which is related to industrial processing of about 2 billion tons of semicrystalline polymeric materials annually as well as protein diseases like Alzheimer's and Parkinson's. Due to connectivity and flexibility of long chain, polymer crystallization involves not only inter-chain position and other orderings but also intra-chain conformational ordering like *gauche*-*trans* or coil-helix transitions, which brings in a new ordering dimension as comparing with spherical "atoms" and small molecules. Unfortunately, although conformational ordering is the most peculiar and critical step in polymer crystallization, how flexible chains transform into conformational ordered rigid segments remains nearly untouched yet. If two-step crystal nucleation is universal for all materials, local structure order or density fluctuation should also occur in polymer crystallization. Considering the peculiar zig-zag or helical conformation in polymer crystal, the shape descriptor defining local structure order should incorporate intra-chain conformational order and rotational symmetries of all neighboring atoms (atoms from the same and different chain segments).

In this work, with full-atom molecular dynamic computer simulation we study crystal nucleation of polyethyle (PE) as the representative of polymers. To define local structure order of PE, we introduce a new shape descriptor or order parameter (named as $O_{CB}$) coupling conformational order and inter-chain order, while the validity of $O_{CB}$ on extracting ordered structure is verified by structure entropy $S_2$ as thermodynamic indicator. Density order parameter is expressed by Voronoi volume $V$. With hexagonal and orthorhombic symmetries of PE crystal as shape matching targets, we define their corresponding $O_{CB}$ and named as $H\text{-}O_{CB}$ and $O\text{-}O_{CB}$ respectively. The simulation details and the definitions of parameters are presented in Methods part. By analyzing the simulation results with the shape descriptor $O_{CB}$ and Voronoi volume $V$, we show that two-step nucleation process indeed occurs in PE. At quiescent condition, local structure ordered clusters matched with hexagonal symmetry $H\text{-}O_{CB}$ forms dynamically first and then the stable orthorhombic nuclei emerges inside the hexagonal clusters, during which local structure order fluctuation plays primary role in assisting nucleation.

## Results

**Nucleation process of PE.** Take the results at 375 K as an example, snapshots at two representative times are shown in Figures 1(a) and (b), respectively (Only $O_{CB}$ structures are displayed). As indicated with cyan colored atoms in Figure 1(a), small clusters with $O_{CB}$ value matching hexagonal symmetry ($H\text{-}O_{CB}$) form stochastically. These hexagonal clusters appear and disappear dynamically, but grow in size with time. When the size of the hexagonal clusters reaching a particular level, orthorhombic structure emerges inside the hexagonal clusters. Figure 1(b) shows that the orthorhombic structures $O\text{-}O_{CB}$ (red) are embedded inside the cyan atoms of hexagonal clusters. These orthorhombic structures are stable and grow continuously with time. Figure 2(c) shows the X-ray diffraction (XRD) data for orthorhombic domain in this work, the peaks at 2θ = 21.2° and 24.2° correspond the (110) and (200) crystal planes, respectively. Due to its short lifetime and small size, the hexagonal clusters found here are not regarded as hexagonal phase of PE as that at elevated pressures. Indeed these *H-$O_{CB}$* clusters only possess conformational and inter-chain rotational orders while no chain parallel or orientation order is required (see SI movie 1) and their density is comparable to that of melt as will be shown later. Thus crystal nucleation of PE is demonstrated to be a two-step process assisted by the coupling between conformational and inter-chain rotational orders, which is further confirmed by simulations at 330 and 350 K.

The evolution of the number of $O_{CB}$ structures is counted to represent the nucleation kinetics, which are depicted in Figures 1 (d)-(f) for temperatures at 330, 350 and 375 K, respectively. At all three temperatures, hexagonal clusters emerges immediately after quenching, while nucleation of orthorhombic structure always takes incubation times, which are about 3.6, 7.5 and 14.0 ns at 330, 350 and 375 K, respectively. The temperature dependence of incubation time indicates that higher supercooling favors nucleation. Oppositely, higher supercooling corresponds to slower growth rate, as shown in Figures 1 (d)-(f). The temperature dependence of nucleation and growth rates observed in the simulations agrees well with experimental observations as well as the prediction of CNT.

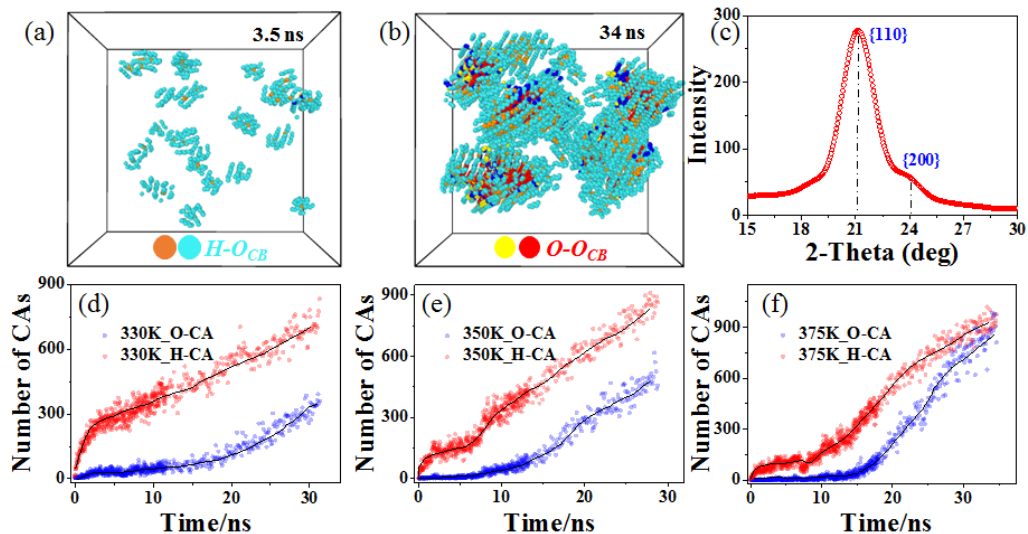

**Figure 1.** (a) and (b) are snapshots of the PE melt nucleation process while different type of atoms are colored differently. Cyan and orange represent $H\text{-}O_{CB}$ clusters, while red and yellow correspond $O\text{-}O_{CB}$ cluster and blue ones belong to both, respectively. (c) The XRD of the orthorhombic clusters. The evolutions of the number of $O_{CB}$ structures at 330 K (d), 350 K (e) and 375 K (f), respectively.

**$H\text{-}O_{CB}$ order independent from density.** To clarify whether $O_{CB}$ order couple with density, Voronoi volume of $O_{CB}$ clusters at different stages are calculated. Figure 2(a) presents a representative of the system at 375 K, where the right column is $O_{CB}$ structures (colored the same as Figure 1) and the left column shows their corresponding Voronoi volume. The emergence of $H\text{-}O_{CB}$ clusters is not accompanied with the change of density as most of them remains in green color (about 30). In SI Figure S2, we show that the densities of the $H\text{-}O_{CB}$ clusters and melt have no difference. Only after the orthorhombic cores form, densification can be distinguished as the color shifts to blue (less than 25) at orthorhombic regions. Note that the density of $H\text{-}O_{CB}$ clusters is relatively low even at the end of simulation. Evidently, $H\text{-}O_{CB}$ and density are two independent orderings, which are not coupled with each other, whilst the formation of orthorhombic structure involves the coupling of density and $O_{CB}$ orders.

*H-O$_{CB}$* clusters does not lead to density fluctuation but does reduce structure entropy ($S_2$), which can provide a practical measure of disorder in the system. For completely disordered systems (i.e., the ideal gas) $S_2$ equals to 0, which becomes negative for ordered structures ($S_2 \to -\infty$ for perfect crystal) (29). To follow the evolution of $S_2$ during the forming and the vanishing process of the *H-O$_{CB}$* clusters, *O$_{CB}$*s at specific time before 7.5 ns are picked out and $S_2$ of these atoms at time range back and forward 100 ps with step of 5 ps are calculated with Equation (1), where *g(r)* is the radial distribution of carbon atoms and $\rho$ is the local density of them within 20 Å. Figure 2(b) presents 5 representatives at 375 K, in which the black dash line marks the formation time of *H-O$_{CB}$* clusters. Irrespective to the formation time, a decline of $S_2$ always occurs during the formation of *H-O$_{CB}$* clusters, which results in a local minimum of $S_2$ (at the shadowed region in Figure 2(b)). We calculate over 20 representatives and they show the same tendency, which supports the shape descriptor *O$_{CB}$* parameter we defined here indeed represents ordered structure with low structure entropy. Combining Figures 2(a) and (b), we can attribute the reduction of $S_2$ to *O$_{CB}$* order rather than density as the later keeps nearly constant in this stage.

$$S_2 = -\frac{\rho}{2} \int_0^{20} dr \{g(r) \times \ln g(r) - [g(r) - 1]\} \qquad (1)$$

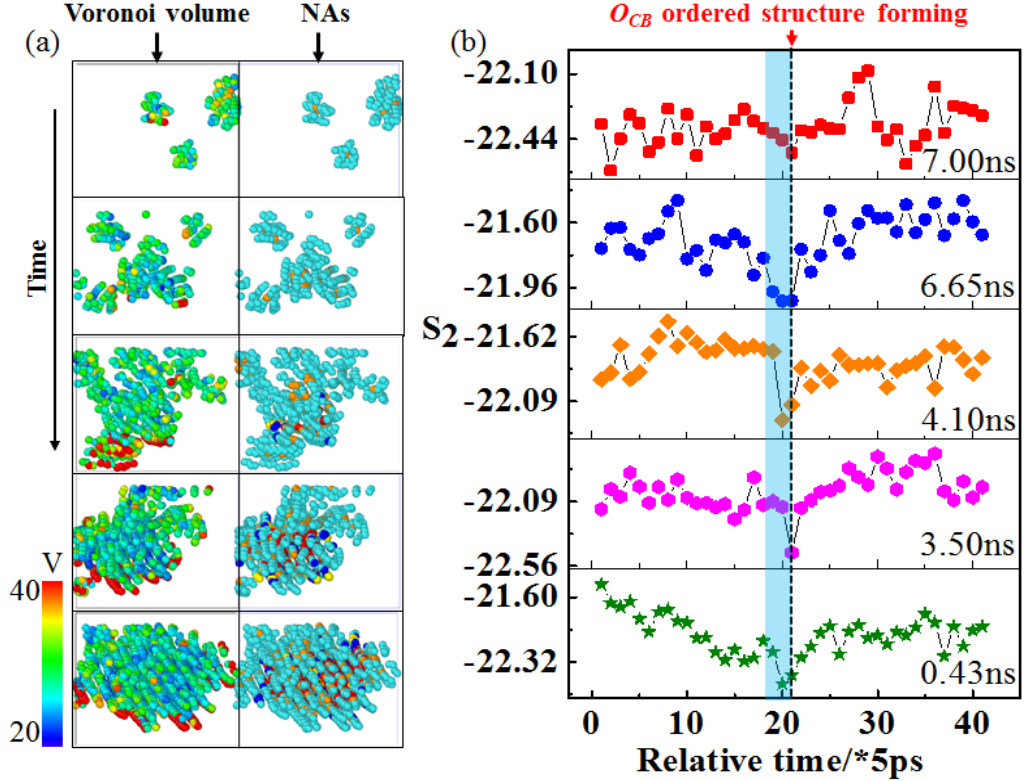

**Figure 2.** (a) $O_{CB}$ structures (right column) and their Voronoi volume of carbon atoms (left column) at 5 representative times, where high Voronoi volume means low density. No correlation between $H$-$O_{CB}$ and density is observed. (b) The $S_2$ evolution of $H$-$O_{CB}$ at 5 representative times before 7.5 ns. The black dash line mark the formation time of $H$-$O_{CB}$ clusters, at which $S_2$ shows a local minimum.

**Coupling of orders and two-step nucleation.** How do $O_{CB}$ and density couple with each other to form stable orthorhombic nuclei? Figure 3(a) depicts the evolution of the number of clusters and the size of the biggest cluster (size are defined as carbon atoms contented in one cluster) in a sub-cube (a crystal grain is contented in it) of the simulation box at 375 K, which can be divided into two-step nucleation and one growth stages. In step I of nucleation (from 0 to 14.0 ns), the number of clusters (red dots with black guide line) increases at the beginning and fluctuates around 6, meanwhile the size of the largest cluster (blue dots with black guide line) grows slowly. Step I is named as $H$-$O_{CB}$ fluctuation because only hexagonal clusters form. The orthorhombic nuclei emerge in step II from about 14.0 to 21.0 ns, which is defined as orthorhombic nucleation step with coupling between $O\text{-}O_{CB}$ and density. In step II, a rapid growth of cluster size is accompanied with a sharp decrease of the cluster number, suggesting that the formation of orthorhombic nuclei proceeds via a coalescence of neighboring $H\text{-}O_{CB}$ clusters. A movie of the simulation process is presented in SI to confirm the coalescence mechanism (see SI Movie 2). Intuitively, merging neighboring $H\text{-}O_{CB}$ clusters to form a large one seems kinetically more favorable than following the stepwise growth process suggested by CNT (1). After the formation of stable orthorhombic nuclei at 20 ns, the number of clusters keeps at a low value and the size of cluster continuously increases, which is the growth stage as illustrated in Figure 3(a).

Accompanied with the formation of orthorhombic nuclei, sharp transitions are also observed on density and $S_2$. We calculate averaged $S_2$ and Voronoi volume of the atoms belong to $O\text{-}O_{CB}$ structures, as these atoms experienced the two-step nucleation and the growth processes, which are plotted vs time in Figure 3(b). In step I of nucleation, the Voronoi volume keeps relatively constant while $S_2$ shows slightly decrease. Once the system enters step II, these two parameters show sharp decreases immediately. The evolution of Voronoi volume supports that $H\text{-}O_{CB}$ is independent with density in step I, while the formation of orthorhombic nuclei in step II does involve coupling between $O_{CB}$ and density.

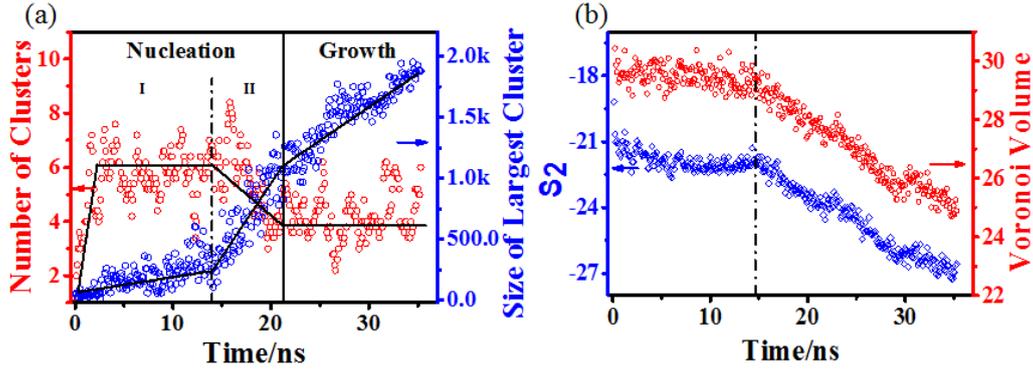

**Figure 3.** (a) The evolutions of the number of ordered clusters and the size of the largest cluster in a representative region of system at 375K. Nucleation of PE is defined into two steps, namely $O_{CB}$ fluctuation step I, orthorhombic nucleation step II. (b) The evolutions of structural entropy $S_2$ and Voronoi volume of carbon atoms at 375K.

The two-step nucleation process is also observed at other two lower temperatures. Figures 4(a) and (b) present the evolutions of the cluster number and the size of the biggest cluster at 330 and 350 K, respectively. At 330 K, step I of nucleation ($H$-$O_{CB}$ fluctuation) terminates at about 3.6 ns, comparatively shorter than 7.5 ns at 350 K and 14.0 ns at 375 K. In step II, the rapid growth of cluster size is also accompanied with the sharp drop of the cluster number at 330 and 350 K, confirming that the formation of stable orthorhombic nuclei follows the coalescence mechanism. Step II takes 7.7, 9.4 and 7.3 ns at 330, 350 and 375 K, which starts (ends) with sizes (carbon atom number) of the biggest clusters of about 180 (800), 250 (1350) and 300 (1100), respectively. Lower temperature corresponds to shorter time of step I and smaller cluster sizes at the onset of step II, while the time and the cluster size at the end of step II do not follow a monotonic trend with temperature, indicating that $H$-$O_{CB}$ fluctuation in step I and the coupling between $O_{CB}$ and density to form orthorhombic nuclei in step II have different temperature dependence. The estimated growth rates of cluster size in step II are 80, 117 and 109 atom/ns at 330, 350 and 375 K, respectively, resembling the dumb-bell shape of nucleation rate along temperature and confirming the validity of the simulation results.

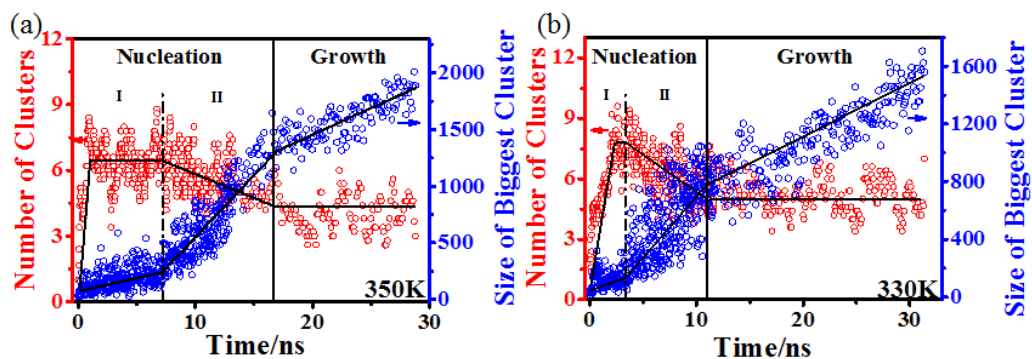

**Figure 4.** The evolutions of the number of ordered clusters and the size of the biggest cluster in a representative region of systems at 350 K (a) and 330 K(b), respectively.

**Discussions and Conclusion**

Based on above analysis on the simulation results with two order parameters, including $O_{CB}$ and density, crystal nucleation of PE is demonstrated to be assisted by the $O_{CB}$ local structure order, which couples between conformational and inter-chain rotational orders, in-line with the two-step scenario observed in colloid systems. The precise kinetic pathway of PE crystal nucleation includes (I) $H$-$O_{CB}$ fluctuation and (II) orthorhombic nucleation. Different from the stepwise growth process suggested by CNT (1), the formation of stable orthorhombic nuclei proceeds via the coalescence mechanism of neighboring $H$-$O_{CB}$ clusters. After the formation of stable orthorhombic nuclei, crystal growth proceeds with increasing both lateral size and thickness. From the movie (see SI movie 2) we notice that crystal grains absorb neighboring $H$-$O_{CB}$ clusters to achieve side expanding and simultaneously grow in chain direction through conformation adjustment, which looks rather similar with the multistage model proposed by Strobl (3) and also found in other systems by De Yoreo et al (30).

Crystallization nucleation is assisted by the $O_{CB}$ local structure order in PE. Intra-chain conformational order is one intrinsic precondition of polymer crystallization, which does not exist in simple spherical atoms. Hoffman-Lauritzen (HL) (28) secondary nucleation model simply assumes that conformational ordered segments with length as the same as lamellar thickness attach on growth front and the lateral surface free energy equals to the entropic loss in conformational ordering of the segments (28). Through the assignment of surface free energy, HL model takes conformational ordering as the rate-limited factor in polymer crystallization. Olmsted et al (24) proposed that coupling between conformational order and density induces liquid-liquid phase separation prior to crystallization. As density fluctuation is global in nature, both concentration and length of conformational ordered segments are required in the model. Focusing on the crystal growth front, the multi-stage crystallization model of Strobl (3) composes of orientational order with long conformational ordered segments though not explicitly stated. Evidently, all these models take either concentration or length of conformational ordered segments as the precondition for preordering or nucleation, while how flexible chains transform into such conformational ordered segments with required length and concentration is not explicitly clarified yet. This challenge may be solved by $O_{CB}$ local structure order as it is defined locally around particles. Our simulations show that short *trans* segments is sufficient to arrange in $H$-$O_{CB}$ ordered clusters without the requirement on orientation (parallel packing) or density order. Indeed without inter-chain coupling, thermal fluctuation may prevent *trans* segments of PE from growing in length and concentration. Here we show that this obstacle is circumvented by the $O_{CB}$ local structure order, which promotes the growths of *trans* segments and hexagonal clusters with low structure entropy $S_2$, although they are still dynamic in nature. After the $H$-$O_{CB}$ clusters reach certain size, the later orderings may set in and eventually result in crystal order. Another interesting observation is the coalescence of neighboring $H$-$O_{CB}$ clusters to form stable orthorhombic nuclei. The existences of bond stretch, angle torsion and conformation increases the free energy barrier for both forming and breaking a nuclei at initial stage. In polymer chain, the connectivity of monomers may indeed prefer to take a collective approach in nucleation and growth, like merging neighboring $H$-$O_{CB}$ clusters.

The occurrence of $O_{CB}$ fluctuation assisted nucleation in PE does not necessarily exclude the possibility of density fluctuation before crystal ordering, especially under non-equilibrium conditions like strong flow (31–33). Density fluctuation, as revealed by small angle X-ray scattering, has been widely reported prior to the onset of crystal order. As suggested by theory and experiment for stretch-induced coil-helix transition (34, 35), flow can also promote the transformation from flexible chain to rigid conformational ordered segments, which may further directly couple with density rather than local orders. Indeed the existence of intra-chain conformational order brings in a far more complex phase behavior than those in spherical atoms. Mutual coupling between two order parameters like intra-chain conformation order and inter-chain $O_{CB}$ or density order may result in isotropic-nematic transition or phase separation, not to say the coupling among multiple order parameters, which may explain the complex phase or self-assembly behaviors with various metastable structures in synthetic and biopolymers.

**Methods**

**Simulation detail.** Full-atom MD simulations are carried out with NAMD (36) to keep conformation and stereo-hindrance effect of PE. The CHARMM force field is chosen with the parameters proposed by Yin and MacKerell et al (37). The system contains 32 PE chains with 500 monomers/chain, so there are about 100,000 atoms in the simulation box. Initial structure of amorphous PE is generated by random walk using Materials Studio packages (38). NPT ensemble is set to control the whole system at pressure of 1 atm. After relaxing 2 ns at 600 K to create PE melt with $<R^2>/<R_g^2> = 5.20 \pm 1.45$ (mean squared end vector $<R^2>$ over radius of gyration $<R_g^2>$), and then the system is quenched to 330 K, 350 K and 375 K to run dynamics over 28 ns, respectively, during which the time step is 1 fs. The periodic boundary condition is imposed in three directions.

**Parameter definition.** Shape descriptor, the mathematical fingerprint to identify a local or global structure, has been widely used in diverse systems (colloidal, protein and nanoparticles) (39). The first step is selecting a reference structure, herein the ideal orthorhombic and hexagonal lattices of PE are built up as reference structures in this work (Figure 5(a), left and right sides, respectively). The key step for a shape descriptor is to construct equations to be a converter transforming the multi-dimension structure into a mathematic index or similarity metric, then the residuals between a query and reference ones can help achieving structure retrieval. Equations (2) and (3) are our 'converters' and Figure 5(b) shows the 'mathematic indexes' for different referential structures named $O_{CB}$, while the $O_{CB}$ value equals to 0.130 corresponding to the Center Atom of orthorhombic cluster (O_CA), while $O_{CB}$ equals to 0.150 or 0.165 for hexagonal structure (H_CA). $Q_l$ in Eq. (2) is summation of several orders of spherical harmonic function $Y_{lm}$, where $l = 4$ and $m \in [0,l]$, $\theta_{ij}$ and $\phi_{ij}$ correspond the polar and azimuthal angles, respectively. Eq. (3) is the average operation, where $N_b(i)$ is the number of neighboring atoms $j$ and $i$ is center atom within a cutoff distance 5.4 Å. We also define $O\_O_{CB}$ and $H\_O_{CB}$ structures as Center Atoms together with their surrounding carbon atoms within distance of 5.8 Å.

$$Q_l = \sum_{m=0}^{l} |Y_{lm}(\theta_{ij}, \phi_{ij})|^2 \qquad (2)$$

$$O_{CB} = \frac{1}{N_b(i)} \sum_{j=1}^{N_b(i)} \left(\frac{2\pi}{l+1} Q_l\right)^{1/2} \qquad (3)$$

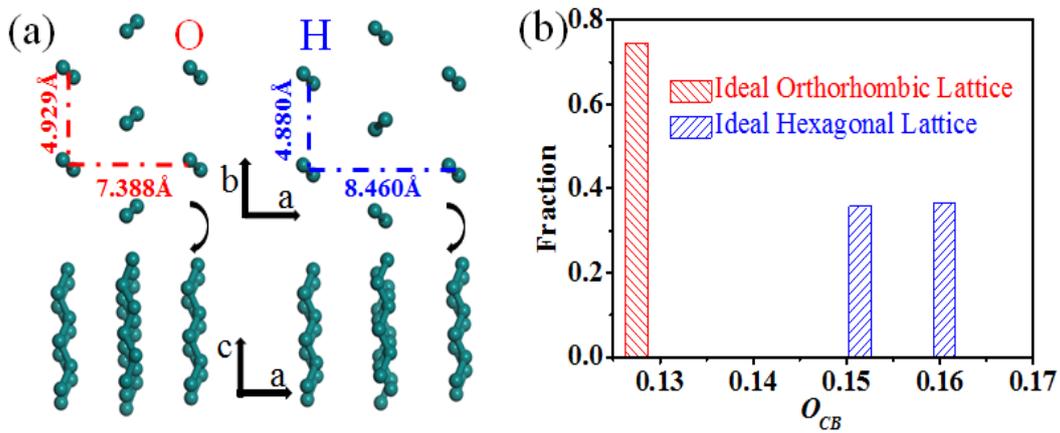

**Figure 5.** (a) Ideal orthorhombic (left) and hexagonal (right) lattices of PE. (c) $O_{CB}$ values of orthorhombic and hexagonal lattices. Orthorhombic structure has $O_{CB}$ equals to 0.130, while $O_{CB}$ for hexagonal lattice equals to 0.150 or 0.165, which are well separated with each other.

**The shape descriptor is constructed based on following considerations:**

a) **Coordinate transform.** One of the principal difficulties faced by our structure retrieval is the isotropic of system, which means the orientation of the chain is unpredictable. The original coordinate would bring catastrophe when adapting it into a 'converter'. Hence the Descartes coordinate was transformed into a self-adaption spherical coordinate. A certain center atom and corresponding polar and azimuthal angles with neighboring atoms within 5.4 Å were calculated. In this way the atom in first shell are contained while the azimuthal angle $\phi$ always in *ab* plane and polar angle $\theta$ expresses the relative position along the chain, which eliminates the effect of isotropic.

b) **Dimension reduction approximation.** The relative spherical coordinate presents the distributions of $\phi$ and $\theta$ and Figure 6(a) shows the essential connections between these two angles for a reference structure that a given $\phi$ must correspond several fixed $\theta$ due to the *trans* conformation. Matching will be difficult for a query structure with ideal crystal especially in polymer system in early stage of nucleation if we restrict both of them when constructing a 'convertor', as the connectivity and the flexibility of chain create many kinds of defects. Herein, only considering the characteristic of $\phi$ or $\theta$ may be a better choice, in terms of effective and realistic.

c) **Characteristic equations**. Eqs. (2) and (3) were constructed to be the 'converter' in this work, whose variables are $\phi$ and $\theta$, which is convenient to the coordinate transformation. The absolute operation of $Y_{lm}$ makes the rotation invariant in *ab* plane (as shown in figure 6(a)), which eliminates the effect of azimuthal angle $\phi$. In Figure 6(b), for a direct comparison the contour of $Q_l$ (green lines) and the position of carbon atom of reference structures (red and blue dots correspond orthorhombic and hexagonal structures, respectively) are moved to the same origin in Descartes coordinate. The features of $\theta$ distribution should be fully considered as the information loss of $\phi$. Figure 6(c) presents the values of $Q_l$ (green dots) and the $\theta$ distributions of reference structures in the same domain ($-\pi/2$, $\pi/2$), their peak positions almost correspond to each other, which indicates the $Q_l$ keeps most of characteristics of $\theta$ distributions and makes it possible to be a proper shape descriptor. Meanwhile, we take conformational order as the precondition because the PE crystals always ask for successive *trans* segments. *Trans* conformation also guarantees the mirror symmetry of $\theta$ distribution and can be regard as a kind of compensate for discarding $\phi$.

d) **Why average**. The sum of squared residuals ($\Delta = \sum_{j=1}^{N_b(i)} |Q_l^{j\_reference} - Q_l^{j\_query}|^2$) should have been the best way to match a query and reference structure based on the value of $Q_l$ if we could know the correspondence of every atom (but we cannot). Alternatively, we take the average of all the neighboring atoms as Eq. (3) did to create reference mathematic indexes (see Figure 5(b)). This operation reduces accuracy remarkably but is sufficient for defining local structure order in PE system, which may be partially due to that conformational order is taken as the precondition. A prediction could be made based on above discussions that slight tilt of neighbor chains will not change a lot of the $O_{CB}$ value as the $\theta$ distribution only goes through a small shifts. Indeed as shown by the simulation result (Figure 6(d) and also SI

Movie 1), local order structures extracted by $H$-$O_{CB}$ can be organized by *trans* segments tilted with each other.

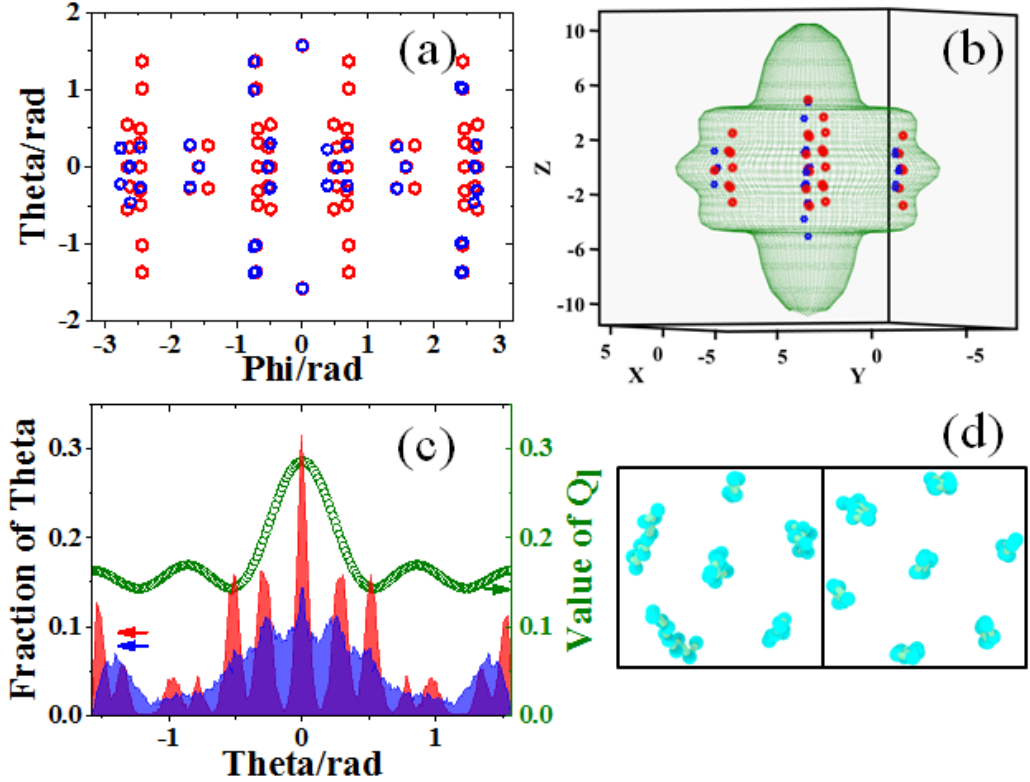

**Figure 6.** (a) The distributions of $\phi$ and $\theta$ for orthorhombic and hexagonal lattices of PE, which indicates the connection between these two angles due to the *trans* conformation and parallel of neighboring segments. (b) The contour of $Q_l$ (green lines), the $Q_l$ have rotate invariant in ab plane. The scatter points are the positions of carbon atoms of reference orthorhombic and hexagonal lattices of PE. (c) The values of $Q_l$ and $\theta$ distributions in same domain ($-\pi/2, \pi/2$), which indicates the $Q_l$ keeps most features of $\theta$ distributions. Red and blue correspond orthorhombic and hexagonal ones in (a), (b) and (c), respectively. (d) The structures we found by $O_{CB}$ parameter.

**Acknowledgements**

The authors would like to thank Prof. Daan Frenkel (Cambridge), Prof. Stephen Cheng (Akron), Prof. Ning Xu (USTC) and Prof. Haojun Liang (USTC) for fruitful discussion on simulation and data interpretation. National Supercomputing Center in Shenzhen, Supercomputing Center of University of Science and Technology of China and National Supercomputer Center in Tianjin (TianHe-1(A)) are acknowledged for providing the computational resources. This work is financially supported by National Natural Science Foundation of China (51633009, 51325301).

Applications to Self Assembly. *Annual Review of Condensed Matter Physics* 2(1):263–285.

**Figure 1.** (a) and (b) are snapshots of the PE melt nucleation process while different type of atoms are colored differently. Cyan and orange represent $H$-$O_{CB}$ clusters, while red and yellow correspond $O$-$O_{CB}$ cluster and blue ones belong to both, respectively. (c) The XRD of the orthorhombic clusters. The time evolutions of the number of $O_{CB}$ structures at 330 K (d), 350 K (e) and 375 K (f), respectively.

**Figure 2.** (a) $O_{CB}$ structures (right column) and their Voronoi volume of carbon atoms (left column) at 5 representative times, where high Voronoi volume means low density. No correlation between $H$-$O_{CB}$ and density is observed. (b) The $S_2$ evolution of $H$-$O_{CB}$ at 5 representative times before 7.5 ns. The black dash line mark the formation time of $H$-$O_{CB}$ clusters, at which $S_2$ shows a local minimum.

**Figure 3.** (a) The evolutions of the number of ordered clusters and the size of the largest cluster in a representative region of system at 375K. Nucleation of PE is defined into two steps, namely $O_{CB}$ fluctuation step I, orthorhombic nucleation step II. (b) The time evolutions of structural entropy $S_2$ and Voronoi volume of carbon atoms at 375K.

**Figure 4.** The evolutions of the number of ordered clusters and the size of the biggest cluster in a representative region of systems at 350 K (a) and 330 K(b), respectively.

**Figure 5.** (a) Ideal orthorhombic (left) and hexagonal (right) lattices of PE. (c) $O_{CB}$ values of orthorhombic and hexagonal lattices. Orthorhombic structure has $O_{CB}$ equals to 0.130, while $O_{CB}$ for hexagonal lattice equals to 0.150 or 0.165, which are well separated with each other.

**Figure 6.** (a) The distributions of $\phi$ and $\theta$ for orthorhombic and hexagonal lattices of PE, which indicates the connection between these two angles due to the *trans* conformation and parallel of neighboring segments. (b) The contour of $Q_l$ (green lines), the $Q_l$ have rotate invariant in ab plane. The scatter points are the positions of carbon atoms of reference orthorhombic and hexagonal lattices of PE. (c) The values of $Q_l$ and $\theta$ distributions in same domain ($-\pi/2,\pi/2$), which indicates the $Q_l$ keeps most features of $\theta$ distributions. Red and blue correspond orthorhombic and hexagonal ones in (a), (b) and (c), respectively. (d) The structures we found by $O_{CB}$ parameter.

# Supporting Information

# Local Structure Order Assisted Two-step Crystal Nucleation in Polyethylene

Xiaoliang Tang[1], Junsheng Yang[1,2], Tingyu Xu[1], Fucheng Tian[1], Chun Xie[1], Liangbin Li[1†]

[1]National Synchrotron Radiation Lab and CAS Key Laboratory of Soft Matter Chemistry, University of Science and Technology of China, Hefei, China

[2]Computational Physics Key Laboratory of Sichuan Province, Yibin University, Yibin, China

## SI-Figure

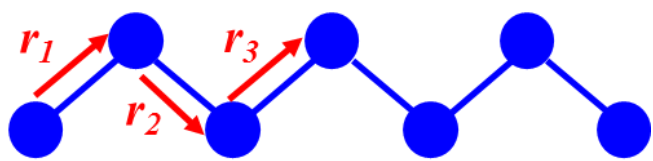

**Figure S1.** The schematic illustration on the definition of conformational order parameter.

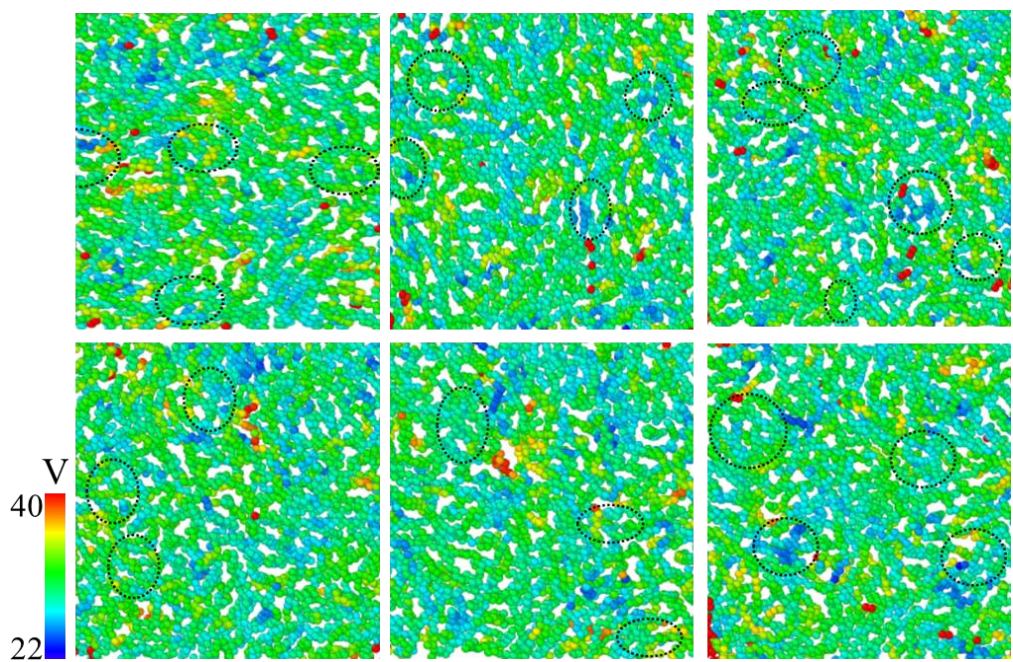

**Figure S2.** The Voronoi volume of carbon atoms in 6 representative snapshots during nucleation. The color bar at left-low corner indicates the volume value. Clusters with hexagonal $O_{CB}$ order is highlighted with the black dash-line circles, which have Voronoi volumes comparable to melt regions.

---

[*] Correspondence author: lbli@ustc.edu.cn

# SI-Text

**Parameter definition**

**Conformational order (*trans* conformation)**

In polyethylene (PE) melt, polymer chains are flexible with statistical random distribution of *gauche* and *trans* conformations, while crystal is packed with conformational ordered segments. Thus melt-crystal transition or crystallization requires the transformation from flexible chain to conformational ordered or *trans* segments, which is the most peculiar and critical step for polymer crystallization, as comparing to spherical atoms or small molecules. Fig. S1 presents an illustration on the definition of *trans* conformational order in this work. With angles of $\overrightarrow{r_1 \times r_2}$ and $\overrightarrow{r_2 \times r_3}$ in the range of 175º to 180º, we regard the two monomers as *trans* segment. As a single *trans* structure is hard to survive under thermal fluctuation, segments with successive *trans* conformations larger than three is considered to be conformational ordered structures.

**Movie 1: Ordered structures**

Several ordered structures are shown in the movie 1, which are extracted out with the $O_{CB}$ order parameters. The first three are taken in step I, which supports that the $O_{CB}$ order parameter does work and may be a better way to distinguish the local order structures of PE or systems with connectivity. The clusters possess intra-chain trans conformational order and inter-chain hexagonal symmetry order locally, but without global orders like orientation (chain parallel) or density (see Figure 5). The shape descriptor we defined in this work allows to distinguish those local order structures with short *trans* segments. As shown in movie 1, in these ordered clusters, short *trans* segments can be tilted to each other, which are largely different from that in ideal hexagonal and orthorhombic crystals but do have a local minimum of structure entropy. The last two structures are taken in the growth step, which have full crystal order.

**Movie 2: Coalescence of hexagonal $O_{CB}$ clusters**

In movie 2, nucleation process in a represent region of the system is presented. An obvious coalescence of neighboring hexagonal $O_{CB}$ clusters to incubate orthorhombic nuclei is observed, which indicates that nucleation of orthorhombic crystals does not follow the standard stepwise growth process as suggested by CNT. In growth stage, crystal grains absorb neighboring hexagonal $O_{CB}$ structures to achieve side expanding and simultaneously grow in chain direction through conformation adjustment, which is similar with the multi-stage model proposed by Strobl (3).